\documentclass[%
 reprint,
]{revtex4-2}

\usepackage{dcolumn}
\usepackage{bm}
\usepackage{graphicx}
\usepackage{subfigure}
\usepackage{graphicx,hyperref}
\expandafter\let\csname equation*\endcsname\relax
\expandafter\let\csname endequation*\endcsname\relax
\usepackage{amsmath}
\usepackage{amssymb}
\usepackage{multirow, makecell, comment} 
\newcommand{\fla}[1]{\begin{flalign}#1\end{flalign}}
\usepackage{braket}
\usepackage{lineno}
\usepackage{lipsum}
\usepackage{xcolor}
\usepackage{soul}
\usepackage{array}

\begin{document}
\preprint{APS/123-QED}
\title{Exponential linewidth narrowing and enhancement of sensitivity in Ramsey interferometry with optically thick ensemble of atoms}

\author{S.A. Moiseev$^1$}
\email{samoi@yandex.ru}
\author{K.I. Gerasimov$^1$} 
\author{M.M. Minnegaliev$^1$}
\author{I.V. Brekotkin$^1$}
\author{E.S. Moiseev$^2$}

\affiliation{ $^1$ Kazan Quantum Center, Kazan National Research Technical University n.a. A.N. Tupolev-KAI, 10 K. Marx St., 420111, Kazan, Russia}
\affiliation{$^2$ Department of Physics, McGill University, 3600 rue University, Montreal, QC, H3A 2T8, Canada}

\date{\today}

\begin{abstract}
Ramsey resonance  is a high-resolution technique used in  spectroscopy, precise measurement of time and frequency, and the creation of modern clocks. 
The Ramsey experiments are typically done in optically dilute samples of atoms to improve homogeneity and avoid back-action of atoms on excitation pulses. 
In contrast to latter belief, we predict and experimentally show that in optically thick resonant media, the back-action can lead to the highly enhanced narrowing of Ramsey resonance. 
The linewidth narrowing and corresponding precision of the frequency measurement scale exponentially with an increase in optical depth of a sample and can reach the limits set by homogeneous broadening. 
We show that this effect is caused by a nonlinear interference of multiple echoes formed inside the atomic medium, which is experimentally confirmed with $^{167}\text{Er}^{3+}$ ions in $\text{Y}_2\text{SiO}_5$ crystal. 
Our findings open new opportunities for sensitivity enhancement of the Ramsey technique, as well as providing researchers with an additional method to investigate fundamental issues of coherent nonlinear interaction of light pulses with resonant atomic ensembles and numerous applications.

\end{abstract}

\maketitle


\textit{Introduction---}
Ramsey's technique \cite{Ramsey1990}  allows for the creation of narrow resonance lines on the atomic and molecular transitions,
which finds application in a wide range of tasks, including high-precision spectroscopy \cite{2023zemach_radius}
detection of weak magnetic fields \cite{bal2012ultrasensitive},
in the development of quantum technologies \cite{2025-NatComm-Hecht},
verification of fundamental physics \cite{dark_matter_searching,RevModPhys.90.045005},
creation of frequency standards and high-precision clocks \cite{Ramsey1957,fartmann2025ramsey}.
The accuracy achieved by recent clock experiments has become high enough to investigate the gravitational redshift \cite{takamoto2020test}.

Ramsey interferometry uses a sample of atoms with two specially selected levels \cite{Ramsey_1949}, which are sequentially illuminated by two resonant electromagnetic pulses with pulse areas of $\pi/2$, separated by a time interval $\tau$ that is less than the lifetime $T_2$ of the quantum coherence of the atomic transition.
After the applications of the pulses, the probability of finding  atoms in the excited state exhibits oscillatory behavior with respect to the detuning $\Delta=\omega_a-\omega_s$, 
with a period of $2\pi/\tau$, known as Ramsey fringes or Ramsey resonances (RRs), which full-width at half-maximum (FWHM)  $\delta\omega_R=2\pi/\tau$ can be much smaller than the spectral width of 
the exciting pulses $\delta\omega_s\sim\delta t_s^{-1}$ ($\delta t_s$ is a temporal duration of light pulse).
Increasing the accuracy of determining the RR frequency (reducing the width $\delta\omega_R$) requires an increase in the delay time $\tau$, and the coherence time $T_2$, respectively.

Ramsey's technique continues to be improved. To narrow the resonance linewidth, Hyper-Ramsey schemes have recently been proposed \cite{Yudin2010, Huntemann2012,2018_ZANON-WILLETTE}, which include a larger number of pulses being tailored with frequency, duration and phase. 
The existing Ramsey technique uses optically thin samples in which atoms do not affect the parameters of exciting $\pi$/2 pulses, while the total amount of atoms increases a signal to noise ratio during population measurement.
Recently, it was pointed out that the transition from optically thin to optically dense resonant media can apparently lead to an increase in the RRs amplitude \cite{Voloshin2019,Voloshin_2022}.
In this Letter, we theoretically predict that the formation of RRs in an optically dense medium of two-level atoms can lead to an additional strong narrowing of RRs caused by the manifestation of a nonlinear resonant interaction of light fields with atoms. 
Namely, the discovered mechanism of narrowing lines provides an exponentially strong decrease in RR linewidth with increasing optical thickness of the atomic resonance  transition.
We also observed the linewidth narrowing with $^{167}\text{Er}^{3+ }$ ions in $\text{Y}_2\text{SiO}_5$ crystal.  
Finally, we discuss the limits of the proposed Ramsey technique, its implementation in prospective media, potential for the future fundamental researches and applications.


\begin{figure}[t]
\includegraphics[width=1.0\linewidth]{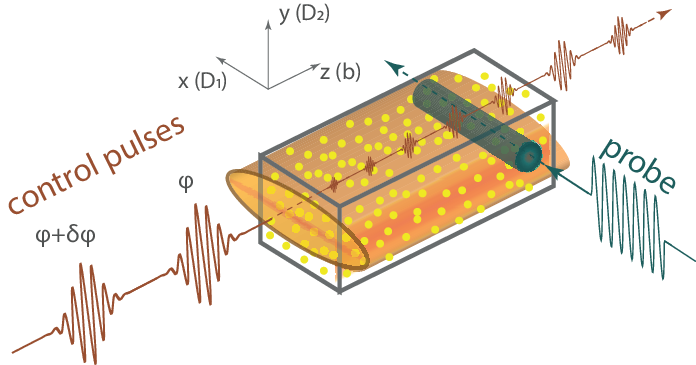}
\caption{Sketch of orthogonal spatial scheme of experiment on RR observation in an optically thick crystal. 
D$_{1}$, D$_{2}$ and b are optical extinction axes of the  Y$_2$SiO$_5$:$^{167}$Er$^{3+}$ crystal. 
A more complete scheme of the experimental implementation is presented in the End Matter.
\label{Fig:scheme} }
\end{figure}

\textit{Mechanism of Ramsey line narrowing---}
The conceptual scheme of the proposed Ramsey interferometry is presented in Fig. \ref{Fig:scheme}. Two laser pulses with pulse areas close to $\pi/2$ are launched into the two-level medium along the same direction on one side of the sample.
The pulse durations $\delta t_{1,2}$ and the time interval between them $\tau$ are set much shorter than $T_2$  and inhomogeneous broadening of of atomic transition $\Delta_{inh}\gg 1/T_2$.
Interaction of light pulses with atoms is described by a semi-classical system of Maxwell-Bloch (also called Maxwell-Bloch and Areccri-Bonifacio) equations \cite{Arecchi1965,McCallHahn1969,Allen_75} (see Supplemental Material (SM) Sec.A \cite{SM} \nocite{McLaughlin1974,HAHN1971,Moiseev1987,Moiseev_JOSAB_25,DeRiedmatten2008,Mims-1968}).

It is convenient to use the pulse area theorem for building intuition on the nonlinear interaction of pulsed light and resonant atoms \cite{McCallHahn1969}:
\fla{
\frac{\partial}{\partial z}\Theta(z)=\frac{\alpha}{2} w(z)\sin \Theta(z),
\label{pulse_area_theorem}
}
where $\Theta(z)=\frac{d}{\hbar}\int \mathcal{E}(t,z)dt$ is a pulse area of light pulse in the point $z$ of the medium, which describes the angle rotation of Bloch-vector of two level atom, $d$ is a dipole moment of atomic transition, $w(z)$ is an inversion of the atomic population ($-1\leq w(z) \leq 1$), $\mathcal{E}(t,z)$ is an electric field envelope,  $\alpha$ is a resonant absorption coefficient \cite{Allen_75}.

The area theorem states that if the pulse area of the input light pulse is greater than $\pi$, then it transforms into an optical soliton inside the medium, acquiring a pulse area of $2\pi$ and propagating unchanged in self-induced transparency (SIT) regime \cite{McCallHahn1969,Allen_75}. 
However, if pulse area of a single pulse is smaller than $\pi$, then pulse is absorbed by the medium (with $\Theta(\alpha z\rightarrow \infty)\rightarrow 0$). 
The bifurcation occurs if the input pulse area is exactly $\pi$. 
In this case, the single $\pi$ pulse can propagate over a large distance inside the medium under conditions of absorption of its energy by increasing its duration, but maintaining the pulse area until perturbations would change the pulse area and bring the pulse into a stable regime.

\begin{figure*}[t]
\includegraphics[width=1.0\linewidth]{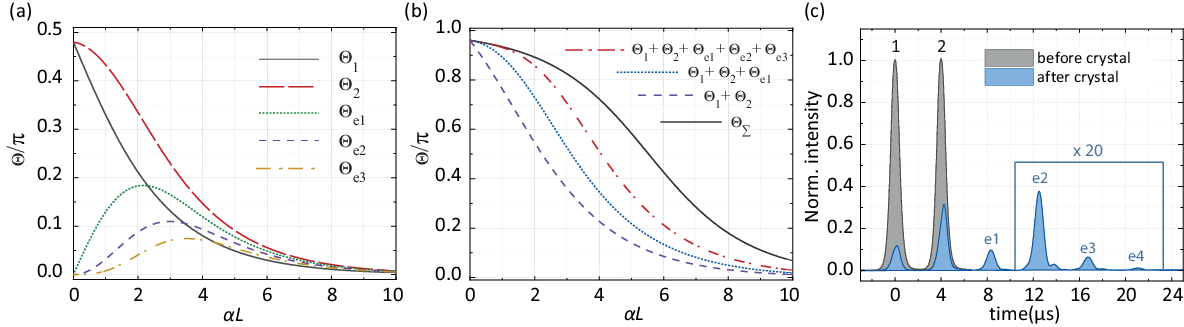}
\caption{(a) Dependence of the pulse areas of exciting laser pulses $\Theta_1 (z)$ , $\Theta_2 (z)$ and photon echo signals $\Theta_{e1} (z)$,  $\Theta_{e2} (z)$, $\Theta_{e3} (z)$  and (b) their various combined values  on the optical depth of the two-level medium for
$\Theta_1 (0) =\Theta_2 (0)=0.48 \pi $.
(c) Four photon echo signals ($e1,e2,e3$ and $e4$)  were observed in experiment in the control beam at the output of the sample having optical depth $\alpha L=2$.
Input signals (1, 2) are shown as grey pulses, unabsorbed part of input pulses and 4 echo signals are depicted as blue pulses.
\label{Fig:Echoes} }
\end{figure*}

Under the condition $\tau\ll T_2$, the ideal ($\frac{\pi}{2}, \frac{\pi}{2}$)  scheme of RRs  formation also occurs at the bifurcation point of Eq.\eqref{pulse_area_theorem}.
However, each of the two input laser pulses is rapidly absorbed at the entrance to the atomic medium ($z\lesssim \alpha^{-1}$) in accordance with Eq.\eqref{pulse_area_theorem}, so that their total pulse area in the depth of the medium evolves as follows $\sim\pi e^{-\alpha z/2}$ (see  Fig. \ref{Fig:Echoes} (a) and  SM Sec.A \cite{SM}). 
However, the total pulse area $\pi$ of the two  input pulses remains unchanged in the depth of the medium  due to the appearance of a cascade of echo signals,  since the two-pulse echo excites the atoms and induces photon echo that  subsequently produces another echo pulse deeper in the medium, and so on.
The pulse areas of the echoes can be extracted with the help of generalization of the pulse area approach \cite{Moiseev2020}. 
For a large time delay ($\tau\gg\delta t_{1,2}$), the atoms are excited in a periodic series of RR lines with an average population inversion close to zero which makes the emission of a Dicke superradiance pulse impossible compared to the excitation of atoms by a single laser $\pi$-pulse \cite{Andreev_1980,2017-Kocharovsky-UFN}. 
Each of the two input pulses and both together are not capable of creating an optical soliton deep in the medium if $\Theta_1(0)+\Theta_2(0)\leq \pi$ \cite{McCallHahn1969} that also simplifies the physical picture and  analysis of the formation of RRs.
Thus the spectral and dynamical properties of the Ramsey technique makes it more stable and reliable in forming ultra-narrow lines comparing with single $\pi$ pulse technique.


The simulation of individual pulse areas of the input two laser pulses ($\Theta_{1},\Theta_{2}$)  and three generated echo signals ($\Theta_{e1}, \Theta_{e2}, \Theta_{e3} $) is shown in Fig.  \ref{Fig:Echoes} (a). 
The choice of the initial pulse areas of $\Theta_1(0)=\Theta_2(0)\equiv\Theta_0= 0.48 \pi$  was due to the limited parameters of our experimental setup (see End Matter).
As can be seen in Fig. \ref{Fig:Echoes} (b), taking into account a larger number of echo signals leads to a more accurate description of the behavior of the total pulse area in the medium describing by Eq.\eqref{pulse_area_theorem}.
After all echo pulses being generated, the  probability of finding atoms in the excited level is calculated using inverse scattering method \cite{Ablowitz_1974} (see SM Sec.B \cite{SM}):

\fla{
P_{bb}(\Delta,z,t)\cong &\frac{e^{-t/T_1}}{1+e^{\alpha z}\cot^2{\Theta_0} }
\frac{1}{\big[1+
\frac{\tan^2{\frac{\Delta \tau+\varphi_{2,1}}{2}} }{\big(\cos^2 {\Theta_0}+ e^{-\alpha z} \sin^2{\Theta_0} \big)}\big]},
\label{P-1}
}

\noindent
where the first and second factors in Eq.\eqref{P-1} determine the maximum probability of $P_{bb}(\Delta,z,t) $ and its spectral shape,
$T_1$ is the lifetime of an optical level ($T_1\gg T_2$), 

\fla{
P_{bb}(\Delta,0,t_{lim})=&\frac{1}{2}\sin^2\Theta_0\big[1+ \cos(\Delta \tau+\varphi_{2,1})\big],
\label{P-2}
}
where $P_{bb}(\Delta,0,t_{lim})$  describes the RRs in an optically thin medium \cite{Ramsey1990},
$t_{lim}$ corresponds to the end of the interaction of atoms with the second laser pulse,
and it is assumed that the intensity of the laser beams is uniform in cross-section,
$\varphi_{2,1}=\varphi_{2}-\varphi_{1}$ ($\varphi_{2}$ and $\varphi_{1}$ are the phases of the laser pulses).
We note that if $\Delta=0$ and $\varphi_{2,1}=0$, the Eq. \eqref{P-1}  coincides with the solution of Eq. \eqref{pulse_area_theorem} for the total pulse area $\Theta(z)$.

The goal of Ramsey interferometry is to determine the detuning $\Delta$ by measuring the population $P_{bb}(\Delta,z,t)$.  
The fundamental sensitivity of the Ramsey experiment on $\Delta$ can be evaluated by estimating Fisher information and  the corresponding Cramer-Rao bound \cite{frieden1998physics}. 
Fisher information for estimating $\Delta$ from population measurement at distance $L$ from the entrance to the medium with a subsequent extraction using a maximum likelihood technique is calculated as   
\fla{
I(\Delta, L) = -\frac{\partial^2 }{\partial \Delta^2} \log P_{bb}(\Delta, L,t).
}

\begin{figure*}
\includegraphics[width=1\linewidth]{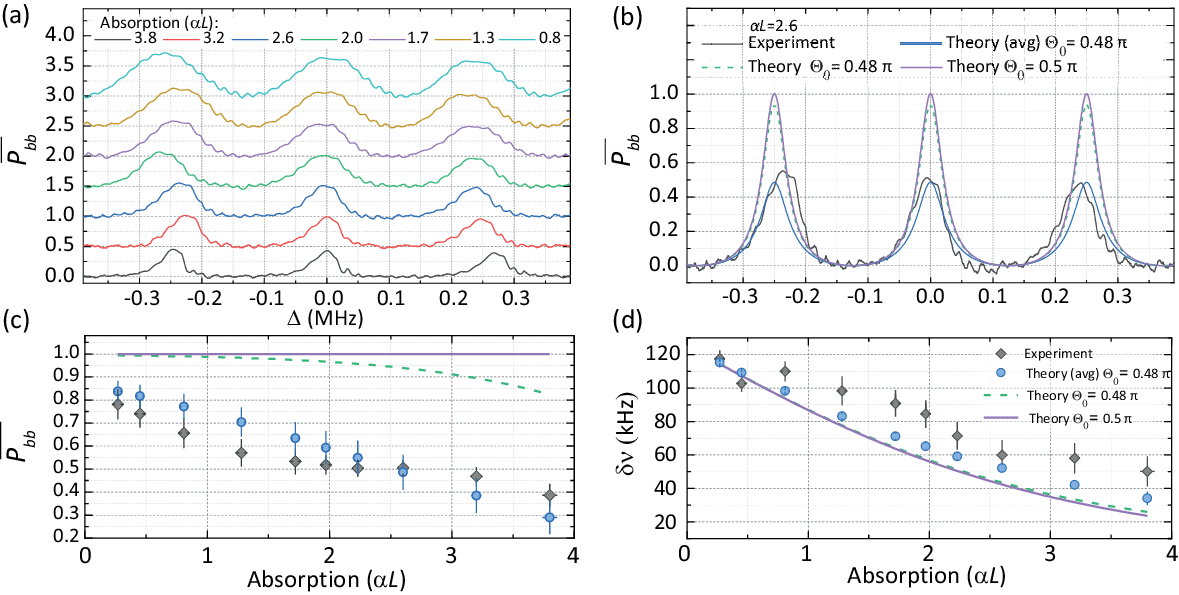}
\caption{(a) Experimental spectra of RR measured for different optical depth of resonant transition  with offset of 0.5 for clarity.  
(b) RR single shot measured for optical depth of 2.6 (black curve), theoretical values for $\Theta_0$=0.48$\pi$ and $\Theta_0$=0.5$\pi$ (green and purple lines, respectively) and theoretical values taking into account the intensity distribution at the intersection of the beams (blue curve) for $\Theta_0$=0.48$\pi$.
(c) Probability of excitation of atoms $P_{bb}$ at $\Delta$ = 0 MHz and (d) peak width $\delta \nu=\frac{1}{2\pi}\delta\omega_R$ (full width at half maximum)   measured for different absorption $\alpha L$ (black diamonds), theoretical values in Eq.\eqref{P-1} for $\Theta_0$=0.48 $\pi$ and $\Theta_0$=0.5 $\pi$ (green and purple lines,respectively) and theoretical values taking into account the intensity distribution at the intersection of the beams (blue circles) for $\Theta_0$=0.48$ \pm 0.01\pi$, $n_x$=1.5 $\pm$ 0.05, $\sigma_{yl}$=1.8 $\pm$ 0.05.
Vertical bars in the experimental data represents $\pm$$\sigma$ uncertainties.
\label{Fig:Ramsey} }
\end{figure*}

Cramer-Rao bound, being an inverse of the Fisher information, defines an uncertainty  (variance) of the parameter estimation around given value. 
For ideal $\pi/2$ excitation pulses ($\Theta_0=\pi/2$) and slow decoherence $T_1, T_2 \gg \tau $, the Cramer-Rao bound of estimating frequency near resonance ($\Delta=0$) becomes (for $\varphi_{2,1}=0$)
\fla{
2\delta \Delta^2 = 2/I(\Delta, L) = 4 e^{-\alpha L}/\tau^2. \label{eq::cramer-rao-bound}
}
Eq.\eqref{eq::cramer-rao-bound} corresponds to FWHM of the RR line ($\delta \omega_R (L)=4e^{-\alpha L/2}/\tau$) at optical depth $\alpha L>2$ when the RR line Eq.\eqref{P-1} takes the form of a Lorentz line, 
where  multiplier $e^{-\alpha L/2}$ describes its exponential narrowing with $\alpha L$ (see also SM Sec.C \cite{SM}).
Atoms with detunings $\Delta=\frac{2\pi n}{\tau}+\Delta'$ with $n=\pm 1, \pm 2,...$ are also excited similarly with the same spectral shape (see experimental data in Fig.\ref{Fig:Ramsey}(a)).


We can summarize the following reasons for the appearance of the described properties of the RR lines.
In the depth of the medium, an increasing number of echo signals are excited, which  is accompanied by a decrease in their amplitude while maintaining the total area of all signals  $\Theta_\Sigma=\pi$ in the ideal ($\frac{\pi}{2}, \frac{\pi}{2}$)    scheme of RRs  formation.
Herein,the spectrum of all echoes remains a period  $2\pi/\tau$, while the linewidth $\delta \omega_R (L)$ of each line shrinks with an increasing number of echoes.
For a single $\pi$ pulse, it is possible to obtain only a single narrowing line however with linewidth wider in $\tau/ \delta t_1 \gg 1$ times than in the case of a two-pulse Ramsey technique.

The linewidth narrowing  reflects an exponential attenuation of the input light energy in the medium, but with a halved absorption coefficient compared to the linear Beer-Lambert law \cite{Beer1852} due to the nonlinearity of the SIT effect.
Accordingly, the number of excited atoms in the RR line decreases exponentially with distance $z$ when measured in a small volume (for $\alpha z\gtrsim 3.5 $):



\fla{
\delta N(z,t)&\approx n_0 S^{1/2} \pi \sigma_{l} ^{2}  \int d\Delta G(\frac{\Delta}{\Delta_{inh}})P_{bb}(\Delta,z,t)= 
\nonumber
\\
& N_0 e^{-(t/T_1+\alpha z/2)},
\label{Delta-N}
}
where  $N_0= n_0 S^{1/2}\sigma_{l} ^{2}\frac{ 2\pi^2}{\tau \Delta_{inh}}$, $n_0$ is an atomic density, $S$ is  the cross-section of the laser beams and $2\sigma_{l}$ is a diameter of the probe light (see SM Sec.E \cite{SM}).

\textit{Experiment ---} 
We test the theoretical prediction by using   
$^{167}\text{Er}^{3+}$ ions doped in  $\text{Y}_2\text{Si}\text{O}_5$ crystal.
Simplified scheme of experiment is shown in Fig.\ref{Fig:scheme}.  
The two control and probe pulses were resonant to the $^4\text{I}_{15/2}-^4\text{I}_{13/2}$ optical transition of $^{167}\text{Er}^{3+}$ ions  (site 2 \cite{2006-Bottger-PRB}) with the following spectroscopical parameters $\Delta_{inh}$=100$\pm$10 MHz, T$_{1opt} \sim$ 10 ms (was measured earlier \cite{Minnegaliev2023}),  T$_M\sim$186 $\mu$s (T$_2$ in the presence of spectral diffusion) was measured in this work by using 2-pulse photon echo experiment at a magnetic field of $\textbf{H}$=3.39 T and a temperature of 1.3 K.
The detailed information about the experimental setup is provided in the End Matter and in SM Sec.F \cite{SM}.




First, we perform two-pulse photon echo measurements for different optical depths. 
An example with an optical depth of $\alpha L\approx 2$ and a control pulse area close to $\pi$/2 is depicted in Fig.\ref{Fig:Echoes} (c).
The figure shows the intensities of the exciting laser pulses, the primary signal, and three subsequent echo signals at the medium output,
where the intensities of  primary echo and  passed two laser pulses match well with the calculated values for their pulse areas, considering $I_{1,2,e1} \propto  |\Theta_{1,2,e1}|^2$,
The intensity of the subsequent three echo signals is noticeably weaker than the primary echo, and their behavior is qualitatively consistent with the theoretical predictions depicted in Fig.2(a), which, in our opinion, is manifested by an increasing 
influence of spatial inhomogeneity of the laser beams in the cross-section, the quantitative description of which will require a separate study elsewhere.


The observation of multiple echoes indicates the possibility of RRs deep in the medium.
The RRs were excited by two control laser pulses with input pulse areas $\Theta_0\approx 0.48 \pi$, temporal duration $\delta t_s=$0.9 $\mu$s, and the delay time between the pulses $\tau=$ 4 $\mu$s. 
Initially, test experiments were carried out, which demonstrated stable control of the spectral parameters of the RRs by controlling the phase difference $\varphi_{2,1}$ between the laser pulses (see SM Sec.F \cite{SM}) with no noticeable influence of phase fluctuations  of these fields.
We then conducted experiments at optical depths ranging from 0.25 to 3.8, the results of which are shown in Fig. \ref{Fig:Ramsey} (a-d).
Figures \ref{Fig:Ramsey} (a,b) show the experimental dependence of the shape and linewidth of the RRs.
Fig. \ref{Fig:Ramsey} (d) presents the experimental data for the behavior of the RR linewidth, which decreased by $\sim 2.4-$times at $\alpha L=3.8$ and was accompanied by a decrease in its amplitude (see Fig. \ref{Fig:Ramsey} (c)).

A comparison of the experimental and theoretical data of the linewidth of the central peak of the fringes $\delta\nu=\delta\omega_{R}/2\pi$ and the average population of the excited level Eq.\eqref{P-1} of atoms $\overline P_{bb}(\Delta, L,t)$ are shown in Fig. \ref{Fig:Ramsey} (b,c), where $t\ll T_{1}$.
The population $\overline P_{bb}(\Delta, L,t)$ was determined from the measurement of the average absorption coefficient $\overline{{\alpha}}_{\perp}(\Delta,L)$ over the x- and y-dimensions of the probe field (see details in End Matter and SM Sec.G \cite{SM}).
In Figs. \ref{Fig:Ramsey} (b-d) together with the experimental data, their theoretical modeling is presented by using Eq.\eqref{P-1} taking into account the influence of the Gaussian intensity profile of the laser beam (the calculations are presented in the SM Sec.G \cite{SM}).  
As can be seen, the theory describes quite well the change of the shape of the RR lines (Fig.\ref{Fig:Ramsey} (b), their amplitude in Fig.\ref{Fig:Ramsey} (c) and linewidth (see Fig.\ref{Fig:Ramsey} (d)).
A slight deviation of the theoretical curve from the experimental data in the description of the RR linewidth (see Fig. \ref{Fig:Ramsey} (d)) may be due to insufficient accuracy in describing the non-uniformity intensity of laser beams and the influence of diffraction effects during their propagation in an optically dense medium.
With an optical depth (OD) of less than 4, it is difficult to distinguish exponential behavior from linear with the existing experimental accuracy and theoretical modeling, but with a larger $OD>4$, an exponential narrowing of the line width becomes noticeable (see Fig. 2 and Fig.7 in SM \cite{SM}).
Therefore, further refinement of the proposed experiment is possible using a larger cross-section of exciting laser beams, which requires increasing the intensity of the master laser.
Using the experimental data and  their theoretical modeling, the spectral sensitivity of the studied RRs is estimated by the Allan deviation $\sigma_a$\cite{wineland1992spin,schulte2020prospects}.
As it is shown in the End Matter, the experimental results demonstrate only a slight improvement of  $\sigma_a$ with an increased optical depth. 
The theory also indicates a possibility of significant (exponential) reduction in $\sigma_a$  by improving the intensity uniformity
of the exciting laser pulses.


\textit{Discussion and conclusion ---} 
A new mechanism for narrowing RR lines has been discovered based on the nonlinear coherent interaction of weak multi-pulse echo radiation with an inhomogeneously broadened resonant medium.
The strong linewidth  narrowing of RRs occurs as a result of the spectral manifestation of the SIT effect when an optically deep coherent resonant medium is excited by several laser pulses.
Recently  a large number of echoes in spin systems on the microwave transitions were observed \cite{Weichselbaumer2020,Debnath2020}.
Based on the above, it can be assumed that in a high-Q resonator (with a linear size $\sim\lambda$) 
multiple spin echoes capable of forming narrowing RR lines, but this requires further research.

Our approach can be extended for Hyper-Ramsey schemes that are promising for suppressing field induced frequency shifts \cite{Yudin2010, Huntemann2012,2018_ZANON-WILLETTE}. These schemes use multiple interrogation pulses with individual pulse being tailored with frequency, duration and  phase. 
The extension of our two pulse method is straightforward, the only condition is to keep the total pulse area of the input light pulses close to $\pi$.
The predicted and experimentally demonstrated mechanism of line narrowing of RRs opens a new direction to solve pressing scientific problems of the interaction of light with coherent resonant media and new practical applications, some of which are noted below.

A wide range of work is carried out in the field of laser physics using optically dense resonant media. 
The proposed Ramsey's technique  will allow us to obtain unique information about the spectroscopic atomic parameters in such media.
These experimental data may be useful for studies of inversion-free lasers \cite{Werren2022,2020-Richter-Optica}, electromagnetically induced transparency and  slow-light control \cite{2025-Chu-PhysRevRes,2025-Hain-NJofPh}, optical Dicke superradiance \cite{Andreev_1980,2017-Kocharovsky-UFN}, optical solitons \cite{McCall1967,alma996752373902959,MAIMISTOV1990} and quantum memory \cite{Lvovsky2009,Tittel2009,Heshami2016a,Chaneliere2018,Guo2023,Zhou2023}.

Of great interest in studying the fundamental laws of the dynamics of the interaction of light fields with optically dense resonant media may be the formation of extremely narrow RRs.
Namely, at high enough OD, e.g. in rare-earth ions doped stoichiometric crystals \cite{Ahlefeldt2016}, the narrowing mechanism Eq.\eqref{eq::cramer-rao-bound} predicts  containing less than one atom in one RR line Eq.\eqref{Delta-N} (see  SM Sec.E \cite{SM}). 
In this case, the continuum macroscopic model of the Maxwell-Bloch equations turns out to be invalid. 
These spectral and spatial effects of the RRs open up new challenges and set requirements for the further development of the theory.
It becomes important to take into account the effect of phase relaxation on the narrowing of the RRs, which requires the development of the inverse scattering method.
This is also true for studying the influence of the quantum properties of light using the quantum generalization of this method \cite{rupasov1982contribution}.



An interesting object is a crystal $\text{Ca}\text{F}_2$ doped with thorium isomers $^{229m}$Th having a resonance line of about 148 nm with a very long lifetime ($\sim$640 seconds) and 
$\Delta_{inh}$ of 4 kHz.
These properties make this crystal promising for studying the implementation of solid-state clocks using the proposed Ramsey technique.
Creating a sample with a high OD of the thorium resonant line is challenging task, but it can be solved by increasing the dopant density.
This method may reduce the lifetime of thorium excited level due to the quenching \cite{Elwell25, Terhune25}. 
However, the resulting negative effect of non-radiative decays can be suppressed by lowering the sample temperature, affecting primarily the level lifetime, rather than the optical depth.
We believe that the media with high optical depth opens up new opportunities for the creation of ultra-precise optical clocks with applications ranging from fundamental physics to high resolution spectroscopy \cite{Peik_2021}.

Thus, the presented results demonstrate the possibility of a significant increase in frequency sensitivity in Ramsey interferometry.
The development of more advanced experimental and measurement methods could provide considerable development of Ramsey interferometry in optically dense media and its potential for solving fundamental problems and numerous applications.

\textit{Acknowledgments ---} 
The authors thank Askhat Basharov for useful discussion.
This research was supported by Ministry of Education and Science of the Russian Federation (Reg.  NIOKTR №: 125012300688-6).
\bibliography{main} 

\onecolumngrid
\begin{center}
    \Large\textbf{End Matter}
\end{center}
\twocolumngrid

\begin{figure*}[t]
\includegraphics[width=0.7\linewidth]{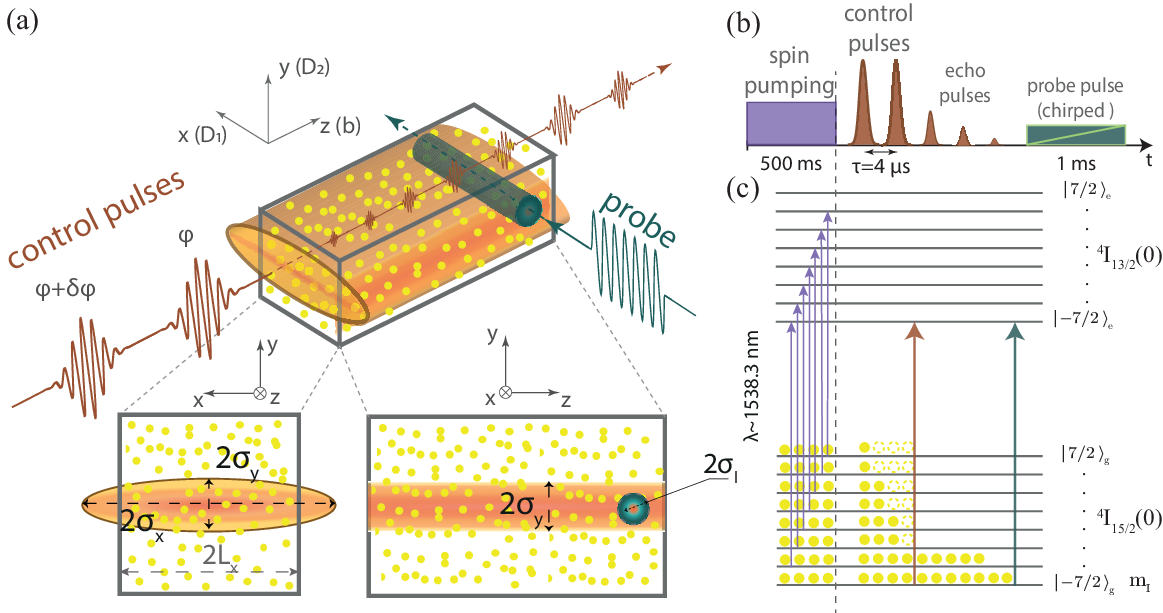}
\caption{(a) Sketch of orthogonal spatial scheme of experiment on RR observation in an optically thick crystal and cross sections ($\sigma_{x}, \sigma_{y}, \sigma_{l}$) of laser beams in two projections. 
D$_{1}$, D$_{2}$ and b are optical extinction axes of the  Y$_2$SiO$_5$:$^{167}$Er$^{3+}$ crystal. 
(b) Time sequence of light pulses. 
Long spin pumping pulse (violet) redistributes the population between hyperfine sublevels of the ground state of erbium ions and  propagates along the z-axis (not shown in part (a) of the figure).
Two control pulses (ocher) generate a series of echo signals creating RRs deep in the medium.
Probe is a frequency chirped weak pulse (green) is used for measuring the absorption spectrum. 
(c) Simplified scheme of optical transitions.
Yellow circles schematically represent the populations of hyperfine sublevels before (left side) and after (right side) spin pumping by the optical technique \cite{2017-Rancic-NP,Minnegaliev2023}.
\label{Fig:schemeEM} }
\end{figure*}

\textit{Experimental details ---}  The two control and probe pulses were resonant to the $^4\text{I}_{15/2}-^4\text{I}_{13/2}$ optical transition of $^{167}\text{Er}^{3+}$ ions in the $\text{Y}_2\text{Si}\text{O}_5$ crystal ($\lambda \sim $ 1538.3 nm.) 
The $^{167}$Er$^{3+}$:Y$_2$SiO$_5$ (c=0.005 \%) crystal had sizes of 3$\times$3$\times$5-mm in the form of  rectangular parallelepiped  crystal with edges parallel to the D$_1\times$ D$_2\times$ b optical extinction axes, respectively (see Fig.\ref{Fig:schemeEM}(a)).
The sources of laser radiation were two tunable single-frequency diode lasers.
The first (master) laser was used to form control and weak frequency chirped probe pulses with linewidth $\leq$3 kHz (250 $\mu$s integration time) in the frequency stabilized regime.
The population was redistributed between hyperfine sublevels of the ground state in erbium ions \cite{2017-Rancic-NP,Minnegaliev2023} by using the second laser frequency of which was swept within about 6 GHz and propagated  along the same path as the control beam.
Time sequence of light pulses and simplified scheme of optical transitions of $^{167}$Er$^{3+}$ ions are shown if Fig.\ref{Fig:schemeEM} (b) and (c), respectively.
The  second laser power was varied to change the fraction of population redistribution and optical depth of the used transition from $\alpha L$=0.25 to $\alpha L$=3.8.

The control beam is passed through a cylindrical lens and has two characteristic sizes: $\sigma_x$ and $\sigma_y$ (see Fig. \ref{Fig:schemeEM}(a)) are the distances from the center of the signal beam in the $x$ and $y$ directions, respectively, at which the beam intensity  decreases by $e^2$ times compared to the central intensity. 
By measuring the beam waists for the control and probe beams by Cinogy CinCam CMOS-1201 IR camera, it was found that $n_x=\frac{\sigma_x}{L_x}$ and $\sigma_{yl}=\frac{\sigma_y}{\sigma_l}$ (used in the numerical simulation) are clamped in the following range: $1.4<n_x<1.55$ and $1.75<\sigma_{yl}<1.85$.
The pulse areas of control pulses of $\Theta_0= (0.48 \pm 0.01)\pi$ are extracted from the theoretical fitting of the experimental data under independently measured constraints for $n_x$  and $\sigma_{yl}$. 
The control pulse amplitudes have been determined to maximize the narrowing of the RR lines with a determination accuracy of $\sim 5-7$\%.
Creating pulses with a pulse area close to 0.5$\pi$ with sufficiently high accuracy will require higher laser  power, as well as more accurate selection and control of the parameters of the cross-sections of laser beams  $n_x$ and $\sigma_{yl}$ (see Fig.4). 
This task is solvable, but requires additional research and resources.


\textit{Accounting for Gaussian intensity distribution of laser beams---}
In the experiment, a spectral absorption coefficient for the weak probe light field propagating along $x-$direction $\alpha_{\perp}(\Delta,x,y,L,t)$ is directly measured to find the probability of excitation $P_{bb}(\Delta,x,y,L,t)$. Based on the relationship between optical density and population difference, it can be seen that:
\fla{
\mathit{\alpha_{\perp}(\Delta,x,y,L)}=\alpha_{\perp}^0 \left[1-2P_{bb}(\Delta,x,y, L)\right],
\label{aL-and-W-1}}
where $\alpha_{\perp}^0 \equiv\mathit{\alpha_{\perp}(P_{bb}=0)}$ and $P_{bb}(\Delta,x,y, L)\cong P_{bb}(\Delta,\Theta_0(r_{\perp}), L)$ (see Eq. (2) of main text where we have replaced $\Theta_0$ by $\Theta_0(r_{\perp})$).

Note that in the case of $\alpha L=3.8$ and $\Theta_0= 0.48\pi$, the shape of the atomic excitation line in Eq. (2) in the main text differs from the Lorentzian (when $\tan^2{\frac{\Delta \tau}{2}}$ is replaced by $\frac{\Delta \tau}{2}$)  by no more than 3.6\%.
The Eq.(2) in the main text 
is  formulated for the plane wave.
For processing the experimental results, it is necessary to adjust the $P_{bb}(\Delta,\Theta_0(r_{\perp}), L)$ by averaging over the spatial intensity distribution because $\Theta \sim E\sim\sqrt{I}$. 
In the experiment, the averaged absorption coefficient $\overline{\alpha}_{\perp}(\Delta,L)$ was determined using exponential attenuation:  
\fla{
\overline {I}(\Delta)=\overline {I}_0(\Delta)\exp\{-2\overline{\alpha}_{\perp}(\Delta,L)L_x\}.
\label{aL_avr}
}

\noindent
$\overline I$ and $\overline I_0$ are the intensities averaged over the $x$ and $y$ coordinates of probe beam at the output and input of the crystal as depicted in Fig. \ref{Fig:schemeEM}, respectively. 
In the SM Sec. G \cite{SM}, we present the calculation of $\overline{\alpha}_{\perp}(\Delta,L)$ and the mean probability of exciting atoms $\overline P_{bb}(\Delta, L)$, which is compared with the experimentally measured shape of the RRs line, its amplitude and linewidth.

\textit{Sensitivity---} 
The properties of spectral sensitivity of Ramsey interferometry can be characterized by analyzing the behavior of the Allan deviation \cite{wineland1992spin,schulte2020prospects}:
\fla{
     \sigma_a =  \frac{\delta \nu }{\sqrt{\delta N}} \sqrt{\frac{T_c}{T_{\text{avg}}}}, 
     }
where $\delta \nu$ and $\delta N$ are the linewidth and number of atoms in the measured resonance line, $T_C$ is a cycle duration and $T_{\text{avg}}$ is averaging time of the measurement.
For sufficiently large optical depth ($\alpha L\gtrsim 3.5$) and the pulse area $\Theta_0\approx \pi/2$, the expressions for the number of atoms and linewidth within the probe region (Eqs. (6) and (5) of the main text) result in the Allan deviation being

\begin{figure*}
\includegraphics[width=1\linewidth]{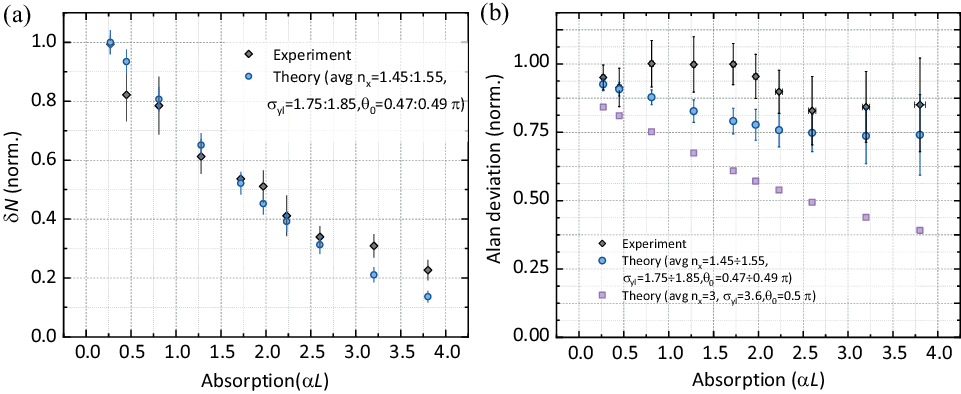}
\caption{
 (a) Normalized number of atoms $\delta N$ under the central RR and (b) Estimate of normalized Allan deviation $\sigma_a$ based on the measured $\delta \nu$ and $\delta N$ for different $\alpha L$ (black diamonds) and its theoretical modeling   for  $\Theta_0$=0.48$ \pm 0.01\pi$, $n_x$=1.5 $\pm$ 0.05, $\sigma_{yl}$=1.8 $\pm$ 0.05 (blue circles) and  for $\Theta_0$=0.5$\pi$, $n_x$=3 , $\sigma_{yl}$=3.6 (magenta squares).Vertical bars in the experimental data represents  $\pm$$\sigma$ uncertainties.
\label{Fig:Alan} }
\end{figure*}

\fla{
\sigma_a =  \frac{2}{\tau} \sqrt{\frac{T_c}{N_0 T_{\text{avg}}}} e^{-\alpha z/4}.
\label{Alan-f}
}
The Eq. \eqref{Alan-f} shows that despite the reduction of the number of atoms, Allan deviation still holds a promise for exponential reduction.
The experimental data for the number of atoms $\delta N$ and the estimates of $\sigma_a$ normalized to its value for a sample with small optical depth ($<0.1$) are shown in Fig. (\ref{Fig:Alan}) along with its theoretical modeling.
As it is seen in Fig. \ref{Fig:Alan} (a), the observed  $\delta N$ is well described by a theory that takes into account inhomogeneity of the intensity of laser beams. 
Experiment shows a slight improvement in the estimated $\sigma_a$ (at $\alpha L>2.5$) as depicted in Fig. \ref{Fig:Alan} (b). We attribute the increased errors to a finite linewidth of the short probe laser pulse that increases error at probing narrow linewidth.   
Fundamentally, increasing the spatial homogeneity of laser beams will significantly improve spectral sensitivity as shown by magenta squares in Fig. \ref{Fig:Alan} (b).

\end{document}